\documentclass[11pt,a4paper]{article}
\usepackage{jcappub}
  
\title{Reconstruction of the HII Galaxy Hubble Diagram using Gaussian Processes}
\author[2]{Manoj~K. Yennapureddy}
\author[1,3]{and Fulvio~Melia%
\note{John Woodruff Simpson Fellow.}}
\affiliation{$^2$Department of Physics, 
              The University of Arizona,
              Tucson, AZ 85721}
\emailAdd{manojy@email.arizona.edu}
\affiliation{$^3$Department of Physics, the Applied Math Program, and Steward Observatory, \\
              \null\hskip 0.05in The University of Arizona, Tucson, AZ 85721}
\emailAdd{fmelia@email.arizona.edu}

\abstract{The Hubble diagram constructed using HII galaxies (HIIGx) and Giant extragalactic 
HII regions (GEHR) as standard candles already extends beyond the current reach of 
Type Ia SNe. A sample of 156 HIIGx and GEHR sources has been used previously
to compare the predictions of $\Lambda$CDM and $R_{\rm h}=ct$, the results of
which suggested that the HIIGx and GEHR sources strongly favour the latter 
over the former. But this analysis was based on the application of parametric 
fits to the data and the use of information criteria, which disfavour the less 
parsimonious models. In this paper, we advance the use of HII sources as 
standard candles by utilizing Gaussian processes (GP) to reconstruct the distance 
modulus representing these data without the need to pre-assume any particular 
model, none of which may in the end actually be the correct cosmology. In addition, 
this approach tightly constrains the $1\sigma$ confidence region of the 
reconstructed function, thus providing a better tool with which to differentiate 
between competing cosmologies. With this approach, we show that the {\it Planck} 
concordance model is in tension with the HII data at more than $2.5\sigma$, while 
$R_{\rm h}=ct$ agrees with the GP reconstruction very well, particularly at 
redshifts $\gtrsim 10^{-3}$. 
}

\begin{document}
\maketitle
 
 \flushbottom

\section{Introduction}
HII galaxies (HIIGx) and Giant extragalactic HII regions (GEHR) share certain
characteristics, including similar optical spectra and massive star formation
\cite{Melnick1987}, and prominent Balmer emission lines in H$\alpha$ and 
H$\beta$ produced by surrounding hydrogen gas ionized by the massive star 
clusters \cite{Searle1972,Bergeron1977,Terlevich1981,Kunth2000}. 
It has been noted that the luminosity $L({\rm H}\beta$) 
in H$\beta$ in these systems is strongly correlated with the ionized gas 
velocity dispersion $\sigma$ \cite{Terlevich1981}, presumably because
both the number of ionizing photons and the turbulent velocity of the gas 
increase with the mass in the starburst component \cite{Melnick2000,Siegel2005}. 
This strong correlation and a relatively small
dispersion in the relationship between $L({\rm H}\beta$) and $\sigma$ makes
it possible to consider and use these galaxies (and HII regions) as standard candles 
\cite{Melnick1987,Melnick1988,Fuentes2000,Melnick2000,Bosch2002,Telles2003,Siegel2005,Bordalo2011,Plionis2011,Mania2012,Chavez2012,Chavez2014,Terlevich2015}.

The first attempt to use HIIGx and GEHR as cosmological tracers was made
by Melnick et al. \cite{Melnick1988}, who used the $L({\rm H}\beta$) - $\sigma$ 
correlation to estimate the Hubble constant $H_0$. Most recently, Terlevich 
et al. \cite{Terlevich2015} optimized the cosmological parameters in the
standard model using 156 sources, including 25 high-$z$ HIIGx, 107 local
HIIGx, and 24 GEHR, demonstrating that these parameter values are
consistent with those obtained through the analysis of Type Ia SNe.

Broadening the reach of the $L({\rm H}\beta$) - $\sigma$ correlation, 
Wei et al. \cite{Wei2016} investigated whether the current sample of HIIGx could be 
used---not only to optimize the parameters of any given model, but also---to 
compare competing models, such as $\Lambda$CDM and the $R_{\rm h}=ct$ universe 
\cite{Melia2003,Melia2007,Melia2013a,Melia2016,Melia2017a,MeliaAbdelqader2009,MeliaShevchuk2012}. 
Using maximum likelihood estimation (MLE), these authors confirmed that the 
parameters in flat $\Lambda$CDM optimized with the HIIGx and GEHR correlation 
(e.g., yielding $\Omega_{\rm m}=0.40^{+0.09}_{-0.09}$, where $\Omega_{\rm m}
\equiv \rho_{\rm m}/\rho_c$ is the normalized matter energy density $\rho_{\rm m}$ 
in terms of the critical density today, $\rho_c\equiv 3c^2H_0^2/8\pi G$) 
are generally consistent with their {\it Planck} values \cite{Ade2014}, though they
went further and argued that the Akaike (AIC), Kullback (KIC) and Bayes 
(BIC) Information Criteria tend to favour the $R_{\rm h}=ct$ model over 
$\Lambda$CDM, with likelihoods $\sim 94.8\%$ versus $\sim 5.2\%$ for AIC,  
$\sim 96.8\%$ versus $\sim 3.2\%$ for KIC, and $\sim 98.8\%$ versus $\sim 
1.2\%$ for BIC. When additional parameters are optimized using this 
approach, the respective probabilities are skewed even more in favour
of $R_{\rm h}=ct$. 

This comparison of $\Lambda$CDM and $R_{\rm h}=ct$ using HIIGx and GEHR
data has confirmed and strengthened similar results obtained previously
with other types of observations. These have included: Type Ia SNe 
\cite{Melia2012,Wei2015b}, based on the framework established
by Perlmutter et al. \cite{Perlmutter1998}, Riess et al. \cite{Riess1998}, and Schmidt et al. \cite{Schmidt1998};
Gamma-ray bursts \cite{Wei2013}, based on correlations and ideas first
proposed by Dai et al. \cite{Dai2004} and Ghirlanda et al. \cite{Ghirlanda2004};  high-$z$ quasars 
\cite{Melia2013b,Melia2014,MeliaMcClintock2015,Kauffmann2000,Wyithe2003}; 
cosmic chronometers \cite{MeliaMaier2013,MeliaMcClintock2015}, 
using an idea first explored by Jimenez \& Loeb \cite{Jimenez2002}
and Simon et al. \cite{Simon2005}; age measurements of passively evolving galaxies 
\cite{Wei2015c}, founded on principles laid out by Alcaniz \& Lima \cite{Alcaniz1999}
and Lima \& Alcaniz \cite{Lima2000}; gravitational lenses \cite{Wei2014,Melia2015,YennapureddyMelia2017}, 
Type Ic superluminous SNe (e.g., ref.~\cite{Inserra2014,Wei2015a}) and the aforementioned
HIIGx and GEHR \cite{Wei2016}. 

In this paper, we advance the use of HIIGx and GEHR as standard
candles in yet another significant and broadly meaningful way. All of the
previous applications of the $L({\rm H}\beta$) - $\sigma$ correlation
have been based on the use of parametric fits inferred from specifically 
chosen models. One of the principal limitations of this approach is that
none of the selected models may actually be the correct cosmology. Here,
we use another statistical method developed recently to reconstruct the
function representing the data without the pre-assumption of any model,
based instead on utilizing Gaussian processes (GP; see, e.g., ref.~\cite{Seikel2012}). 
The GP method has the flexibility of reconstructing a function 
that best fits the data without the assumption of any parametric form
at all, thereby making model comparisons more reliable and robust. 

\section{Observational Data and Model Comparisons} 
The data sample used in this paper consists of 25 high-$z$ HII 
galaxies, 107 local HII galaxies, and 24 giant extra galactic HII 
regions, compromising 156 sources in total (Terlevich et al. 2015). 
Of these 156 sources, the 107 local HII galaxies lie in the redshift 
range $0.01\lesssim z \lesssim 0.2$ \cite{Chavez2014}. 

For these sources, the luminosity versus ionized gas velocity correlation 
\cite{Chavez2012,Chavez2014,Terlevich2015} is given as
\begin{equation}
\log L({\rm H}\beta)=\alpha \log \sigma ({\rm H}\beta) +\kappa\;,
\end{equation}
where $\sigma$ is the velocity dispersion of the H$\beta$ line, and
$\alpha$ and $\kappa$ are constants representing the slope and intercept. 
In general, the values of these parameters have to be optimized simultaneously 
with the cosmological parameters to avoid circularity issues, but they appear
to be very insensitive to the underlying cosmology. Wei et al. \cite{Wei2016} 
carried out this optimization procedure by combining $\kappa$ and the Hubble 
constant $H_0$ together as 
\begin{equation}
\delta =-2.5\kappa-5\log H_0 +125.2\;,
\end{equation}
and found that regardless of which model one assumes, $\alpha$ and $\delta$
deviate by at most only a tiny fraction of their $1\sigma$ errors from one
application to the next. For simplicity, we shall therefore use the averages
of the optimized values reported by Wei et al. \cite{Wei2016}, which are
$\alpha=4.87^{+0.11}_{-0.08}$ and $\delta=32.42^{+0.42}_{-0.33}$. To give
a concrete sense of what this means in practice, one would get $\alpha=4.86^{+0.08}_{-0.07}$ 
and $\delta=32.38^{+0.29}_{-0.29}$ for $R_{\rm h}=ct$ and $\alpha=4.89^{+0.09}_{-0.09}$ 
and $\delta=32.49^{+0.35}_{-0.35}$ for $\Lambda$CDM, if one were to re-optimize them 
separately. In the spirit of what we are trying to do here, i.e., reconstructing the 
correlation function independently of any model, we shall therefore adopt the average 
values for all the cases we consider.

Using the $L({\rm H}\beta$) - $\sigma$ correlation, one can obtain the distance 
modulus for an HII galaxy as 
\begin{equation}
\mu_{\rm obs}=-\delta +2.5 \big[\alpha \log \sigma({\rm H}\beta)-\log F({\rm H}\beta)\big]\;.
\end{equation}
By comparison, the theoretical distance modulus is given as 
\begin{equation}
\mu_{\rm th}=5\log\bigg[\frac{\tilde{D}_L(z)}{Mpc}\bigg]\;,
\end{equation}
where $\tilde{D}_L(z)$ is expressed as $\tilde{D}_L(z)=H_0 D_L(z)$,
in terms of the Hubble constant $H_0$ and the luminosity distance $D_L(z)$, 
which depends on the cosmological model. The expressions for $D_L^{\Lambda{\rm CDM}}(z)$
and $D_L^{R_{\rm h}=ct}(z)$ are given in ref.~\cite{Wei2016}.
These two luminosity distances are indistinguishable at low redshift, but
deviate from each other progressively with increasing $z$, as one may see
in figure~1, which shows the distance modulus predicted by Equation~(2.4).

\begin{figure}
\begin{center}
\includegraphics[width=0.7\linewidth]{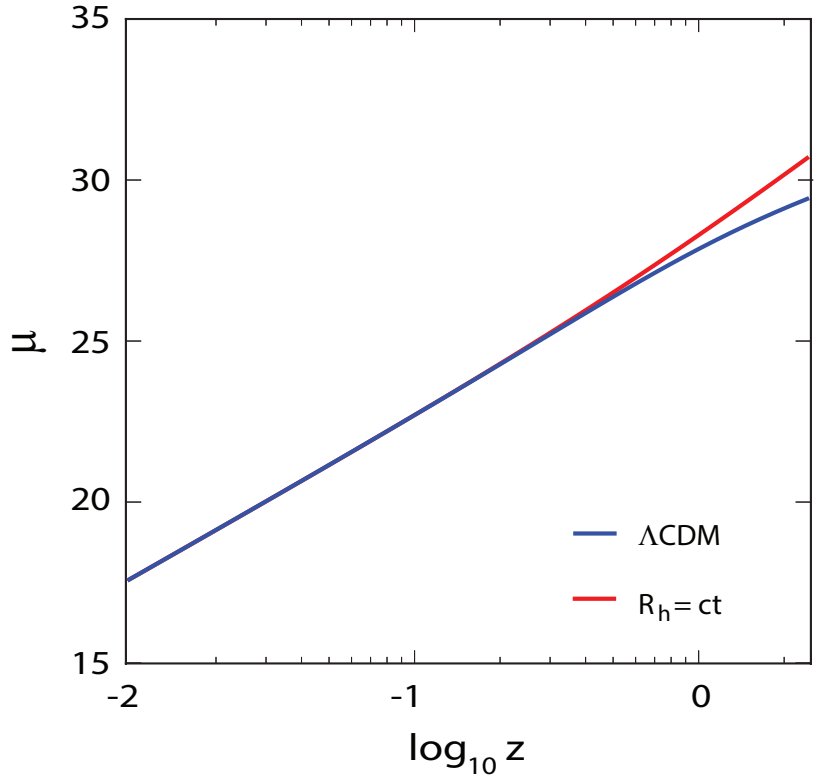}
\end{center}
\caption{The theoretically predicted distance modulus
as a function of redshift for $R_{\rm h}=ct$ (red) and the concordance
$\Lambda$CDM model (blue). While the two models are indistinguishable for
$z\lesssim 0.3$, they deviate significantly at higher redshifts.}
\end{figure}

\begin{figure}
\begin{center}
\includegraphics[width=0.7\linewidth]{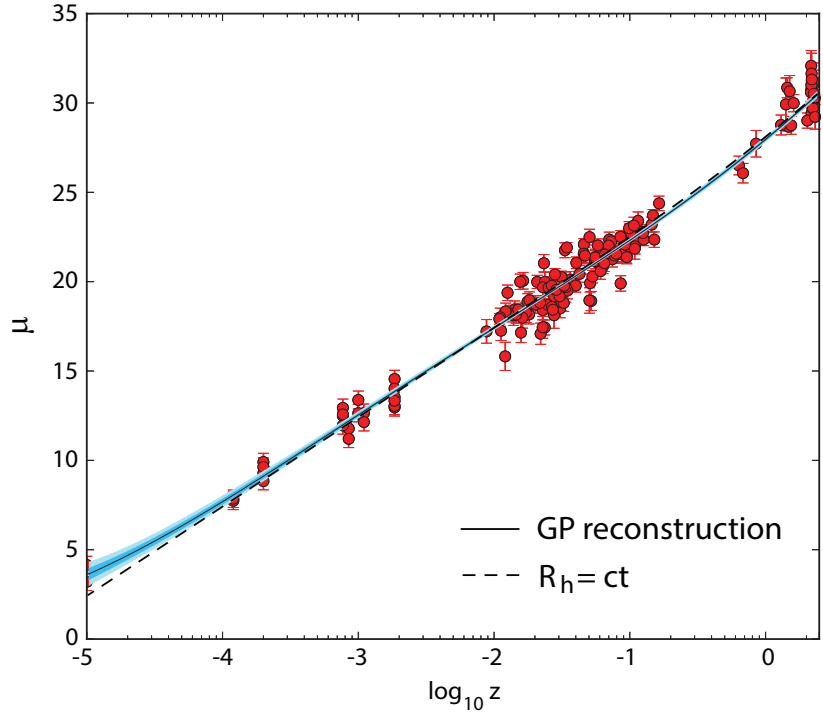}
\end{center}
\caption{The thin solid curve (black) indicates the
reconstructed distance modulus $\mu_{\rm obs}(z)$, using Gaussian processes,
for all 156 of the currently available HII-region/Galaxy data, 
shown as red circles with $1\sigma$ error bars. The dark blue swath
represents the $1\sigma$ confidence region, while light blue is 
$2\sigma$. Also shown in this figure is the distance modulus predicted
by the $R_{\rm h}=ct$ universe (dashed). The paucity of data at very low 
redshifts leads to a less precise determination of the measured 
$\mu_{\rm obs}(z)$, indicated by the widening $1\sigma$ and $2\sigma$ 
confidence regions, but the comparison between the two curves for 
$z>10^{-4}$ is excellent.}
\end{figure}

\section{Gaussian Processes and the HII Galaxy Hubble diagram}
Adapting the GP code developed by Seikel et al. \cite{Seikel2012}, we reconstruct 
the function $\mu_{\rm obs}$ in Equation~(2.3) representing the 156 HII galaxy
sources in Terlevich et al. \cite{Terlevich2015}, without pre-assuming any model or its
parametric form of the distance modulus. The GP method uses some of the attributes of a 
Gaussian distribution, though the former utilizes a distribution over functions obtained 
using GP, while the latter represents a random variable. The reconstruction of a function 
$f(x)$ at $x$ using GP creates a Gaussian random variable with mean $\mu(x)$ and variance 
$\sigma(x)$. The function reconstructed at $x$ using GP, however, is not independent of 
that reconstructed at $\tilde{x}=(x+dx)$, these being related by a covariance function 
$k(x,\tilde{x})$. Although one can use many possible forms of $k$, we use one that 
depends on the distance between $x$ and $\tilde{x}$, i.e., the squared exponential 
covariance function defined as
\begin{equation}
k(x,\tilde{x})=\sigma_f^2\exp\left(-\frac{(x-\tilde{x})^2}{2\Delta^2}\right)\;.
\end{equation}
Note that this function depends on two hyperparameters, $\sigma_f$ and $\Delta$,
where $\sigma_f$ indicates a change in the $y$-direction and $\Delta$ represents a 
distance over which a significant change in the $x$-direction occurs. Overall, these 
two hyperparameters characterize the smoothness of the function $k$, and are trained 
on the data using a maximum likelihood procedure, which leads to the reconstructed 
$\mu_{\rm obs}$ function in Equation~(2.3).

One of the principal features of the GP approach that we highlight in this application
to the HII Galaxy Hubble diagram concerns the estimation of the $1\sigma$ confidence 
region attached to the reconstructed  $\mu_{\rm obs}$ function. The 
$1\sigma$ confidence region depends on both the actual errors of individual data
points, $\sigma_{\mu_{\rm obs}}$, on the optimized hyperparameter $\sigma_f$,
and on the product $K_*K^{-1}K_*^T$ (see ref.~\cite{Seikel2012}), where 
$K_*$ is the covariance matrix at the point of estimation $x_*$, calculated 
using the given data at $x_i$, according to 
\begin{equation}
K_*=[k(x_1,x_*),k(x_2,x_*),...,k(x_i,x_*)]\;.
\end{equation}
$K$ is the covariance matrix for the original dataset. Note that the dispersion at point 
$x_i$ will be less than $\sigma_{\mu_{\rm obs}}$ when $K_*K^{-1}K_*^T>\sigma_f$, i.e.,
when for that point of estimation there is a large correlation between the data. From the above  
Equation  it is clear that the correlation between any two points $x$ and $\tilde{x}$ 
will be large only when $x-\tilde{x}<\sqrt{2}\Delta$. This condition, however, is 
satisfied most frequently for the HII Galaxy data used in our study, which results
in a GP estimated $1\sigma$ confidence region that is smaller than the errors in 
the original data. Since the $\sigma_{\mu_{\rm obs}}$ of the
data are relatively small, the reconstructed 1$\sigma$ regions are correspondingly
small as well, as one may see in figures~2 and 3. This feature is critical to 
understanding why the GP approach for reconstructing the $L({\rm H}\beta$) - 
$\sigma$ correlation yields a powerful diagnostic for comparing competing models. 
We refer the reader to ref.~\cite{Seikel2012} for further details.

\section{Discussion}
The reconstructed $\mu_{\rm obs}$ function (solid), along with its $1\sigma$ 
(dark blue), and $2\sigma$ (light blue) confidence regions, is shown in figures~2 
and 3, together with the 156 sources (with errors) used to derive it. In these figures,
we also show the theoretically predicted distance modulus (dashed) for $R_{\rm h}=ct$
(fig.~2) and $\Lambda$CDM (fig.~3). It is quite evident that the distance modulus
predicted by $R_{\rm h}=ct$ is an excellent match to the reconstructed $\mu_{\rm obs}(z)$
function, while the prediction of $\Lambda$CDM is an equally good fit at low
redshifts, but a very poor fit for $z\gtrsim 0.3$. 

\begin{figure}
\begin{center}
\includegraphics[width=0.7\linewidth]{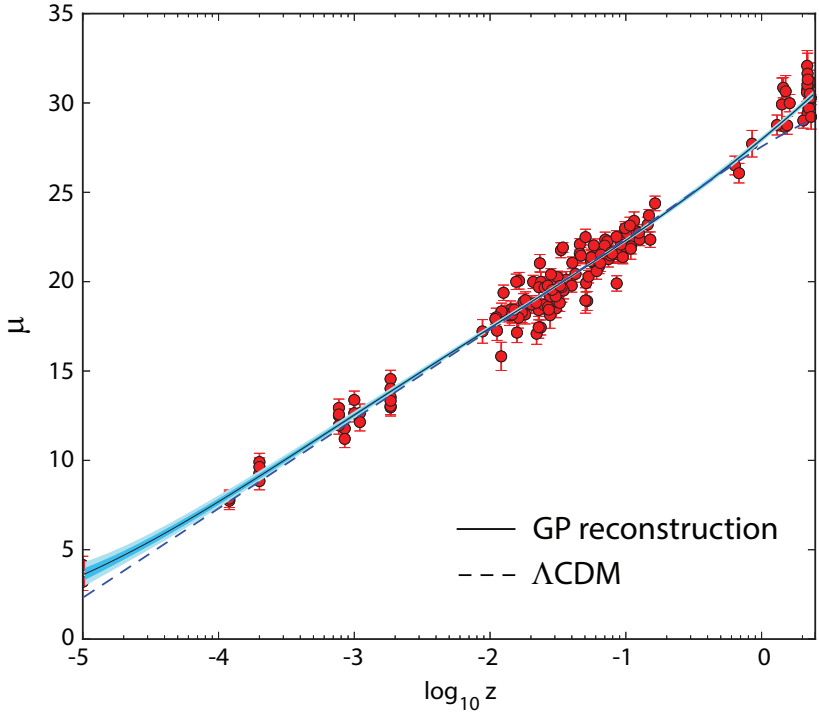}
\end{center}
\caption{Same as fig.~2, except now showing the GP reconstructed
$\mu_{\rm obs}(z)$ curve in comparison with the prediction of $\Lambda$CDM.
Unlike $R_{\rm h}=ct$, the standard model's prediction deviates significantly 
from the GP reconstructed $\mu_{\rm obs}(z)$ curve at $z\gtrsim 0.1$, as 
quantified by the cumulative probability distribution shown in fig.~4 below, 
based on the area differential between the two curves.}
\end{figure}

As we are comparing two continuous functions i.e., $\mu_{\rm obs}(z)$ with either 
$\mu_{R_h=ct}$ or $\mu_{\Lambda CDM}$, we estimate each model's probability of being 
correct using an area minimization statistic used more commonly in other fields, e.g., 
medical diagnostics (see ref.~\cite{DeLong1988}). In this approach, we create 500 
mock samples of HII data according to  
\begin{equation}
\mu_i(z)=\mu_{\rm obs}(z)+r\sigma_{\mu_{\rm obs}}\;,
\end{equation}
where $\mu_{\rm obs}$ are the actual measurements shown in figures~2 and 3 and 
$\sigma_{\mu_{\rm obs}}$ are their associated errors, while $r$ is assumed to be a
Gaussian random variable with zero mean and a variance of $1$. Then we use $\mu_i(z)$ 
together with the $\sigma_{\mu_{\rm obs}}$ errors to reconstruct the function 
$\mu_{\rm mock}(z)$ corresponding to each mock sample, and finally we calculate the 
normalized absolute area difference between $\mu_{\rm mock}(z)$ and the GP reconstructed 
function of the actual data using 
\begin{equation}
\Delta A=\frac{\int_{z_{min}}^{z_{max}}dz^\prime\; \bigg|\mu_{\rm mock}(z^\prime)-
\mu_{\rm obs}(z^\prime)\bigg|}{\int_{z_{min}}^{z_{max}}dz^\prime\;\mu_{\rm obs}(z^\prime)}\;.
\end{equation}

We repeat this procedure 500 times to build a distribution of frequency versus 
area differential $\Delta A$, and from it construct the cumulative probability 
distribution shown in fig.~4. Also shown in this figure is the 
normalized area differential ($0.0018$) for $R_{\rm h}=ct$, corresponding
to a probability of $\sim 69\%$. A close inspection of figure~2 reveals that
the main contribution to this area differential comes from the very
low-$z$ region, where a relative paucity of sources mitigates the accuracy with
which the $\mu_{\rm obs}(z)$ function may be reconstructed. By comparison, the
normalized area differential for $\Lambda$CDM is $0.02$, corresponding to 
a probability $\lesssim 1\%$. For this model, most of the contribution to the 
area differential comes from the high-$z$ measurements, which are of principal 
interest to our model comparison in this paper.

\begin{figure}
\begin{center}
\includegraphics[width=0.7\linewidth]{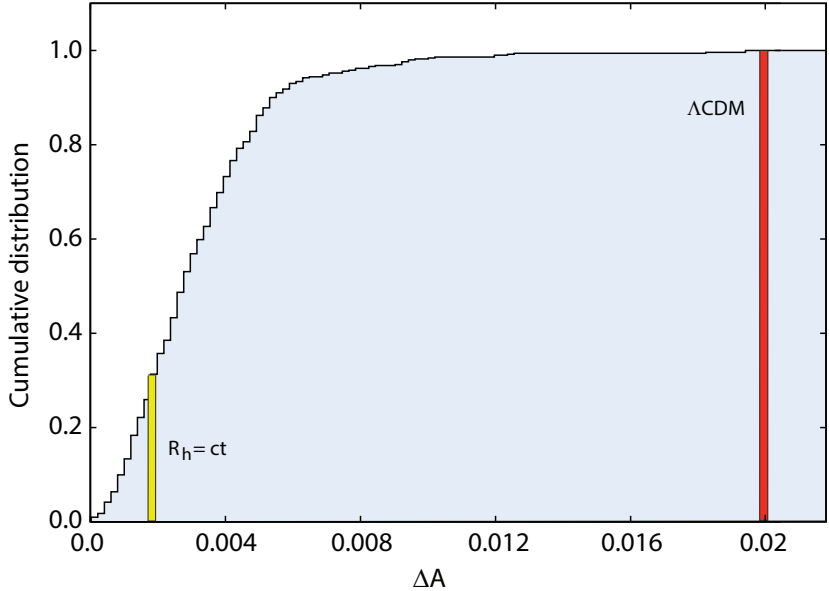}
\end{center}
\caption{Cumulative probability distribution (normalized to 
$1$) of the differential area calculated for $\mu(z)$, based on mock samples
constructed via Gaussian randomization of the measured $\mu(z_i)$ values
(see figs.~2 and 3). The $R_{\rm h}=ct$ model is shown with a yellow bar, corresponding
to an area differential of $0.0018$, with a probability $\sim 69\%$. The $\Lambda$CDM
model is shown with a red bar, corresponding to an area differential of $0.02$,
with a probability $\lesssim 1\%$.}
\end{figure}

As discussed in \S~3 above, one of the strengths of the GP method
for reconstructing the function representing the data is the 
relatively small uncertainty in the result, as illustrated by
the rather tight $1\sigma$ and $2\sigma$ bands shown in figs.~2 and 3. 
This is what allows a comparison of this curve with the predictions
of models being tested to select the preferred cosmology with a 
high degree of confidence. The cumulative probability plot in
figure~4 shows that the predicted fit in $R_{\rm h}=ct$ is
closer to the GP reconstruction than that of $69\%$ of all
the mock realizations, while the best-fit curve in $\Lambda$CDM
does worse than $\sim 99\%$ of the possible variations. Another
way to think of this is that, while the prediction of $R_{\rm h}=ct$ 
lies within about $1\sigma$ of the reconstructed curve, $\Lambda$CDM
is excluded at over $2.6\sigma$.

This outcome refutes some of the conclusions drawn,
e.g., in ref.~\cite{Bilicki2012}, where the analysis was also
based on the use of Gaussian Processes to reconstruct an observed
function. But there are actually two issues to discuss in any
such comparison. These authors did indeed also use a GP approach 
for the analysis of redshift-dependent data, but their focus was
on the Hubble constant $H(z)$. They reconstructed the expansion 
rate, as given by $H$, whereas we are using the HII galaxy Hubble 
diagram, which provides the distance modulus as a function of
redshift. These are two very different kinds of observation. 
Ultimately they must be related, of course, but our reconstructed 
function is of a redshift-dependent distance, while theirs
is of the redshift-dependent expansion rate.

The second issue has to do with which model is favoured by 
the data. Their analysis is now quite dated. Updates to that work 
have appeared in refs.~\cite{MeliaMaier2013,MeliaMcClintock2015,Wei2017}.
A central question is which data ought to be used for the $H(z)$ 
analysis. As we now understand, mixing measurements of $H(z)$ 
based on BAO measurements with those extracted from cosmic 
chronometers, as was done in this earlier work, introduces 
significant biases and systematic uncertainties. The limitation
of BAO measurements is that one must pre-assume a particular model 
in order to disentangle the effects of redshift-space distortions 
from the actual cosmological redshift variation across a
galaxy cluster. Most of the BAO measurements rely on the use 
of $\Lambda$CDM to carry out this step, making the data, 
including $H(z)$, model-dependent. In order to carry out a truly
model-independent comparison between different models based on 
$H(z)$, one must therefore restrict the analysis solely to the 
cosmic chronometer measurements. And when one does this, the 
outcome of the model selection changes from $\Lambda$CDM to 
$R_{\bf h}=ct$, as described more fully in these subsequent
publications.

As we say, however, the present paper does not deal with 
$H(z)$ measurements, so this broader discussion concerning which 
model is favoured by the expansion-rate measurements does not
apply here, though our results in this paper are fully consistent
with these earlier model comparisons based on the Hubble constant $H(z)$ . 
As far as we know, the present paper presents the first application 
of GP processes to the Hubble diagram constructed from HII galaxies.

\section{Conclusion}
Broadly speaking, the results of our analysis in this paper
confirm and strengthen the earlier conclusion drawn from a comparison
of $R_{\rm h}=ct$ and $\Lambda$CMD based on a maximization of the 
likelihood function using the predicted distance modulus in these 
models to fit the data \cite{Wei2016}. That analysis took into 
account the different number of model parameters to arrive at a 
relative probability estimated using various information criteria. 

The analysis of HIIGx and GEHR data with Gaussian processes, which
we have used exclusively in this paper, constitutes a powerful
complementary diagnostic for several reasons. First, there is no need
to pre-assume a parametric form of the distance modulus, based on
specific cosmological models, in order to analyze the data. The
principal benefit of this approach is that, in the end, neither of 
the models being tested may be the correct cosmology, so the GP 
reconstructed function is an overall better representation of 
the data than what either model can provide. Second, the manner 
with which the GP approach identifies the functions that fairly 
represent the data produces a $1\sigma$ confidence region 
significantly tighter than the error one may infer on the basis 
of a simple $\chi^2$ or MLE evaluation. This is the reason why 
the reconstructed function in figures~2 and 3, and its associated 
$1\sigma$ and $2\sigma$ confidence regions, can differentiate 
between models such as $R_{\rm h}=ct$ and $\Lambda$CDM so powerfully, 
as quantified in figure~4. 

On the basis of our analysis, $\Lambda$CDM does not fare very
well when compared to the reconstructed $\mu_{\rm obs}(z)$ function,
irrespective of whether or not it is being compared to $R_{\rm h}=ct$.
Even if we were to consider its predicted distance modulus directly
in the context of the cumulative distribution (fig.~4) constructed 
from the data, the resultant probability is smaller than $\sim 2.5\sigma$.
By comparison, the prediction of $R_{\rm h}=ct$ is virtually 
indistinguishable from the reconstructed $\mu_{\rm obs}(z)$ function,
deviating from it only slightly at very low redshifts, where a 
relative paucity of data weakens the precision with which the
reconstruction may be made. Quite tellingly, the reconstructed
function strongly prefers $R_{\rm h}=ct$ over $\Lambda$CDM
at high redshifts, where the cosmological model comparison is
most important.

Through the use of Gaussian processes in this paper, we 
have strengthened the case for using the $L({\rm H}\beta$) - $\sigma$ 
correlation seen in HIIGx and GEHR data as a standard candle for
testing cosmological models even beyond the current reach 
(i.e., $z\lesssim 1.8$) of Type Ia SNe. This argument will
become even more compelling as the systematic uncertainties in
the $L({\rm H}\beta$) - $\sigma$ correlation are better understood.
These uncertainties include the size of the burst, the age of the burst, 
the oxygen abundance of HIIGx, and the internal extinction correction 
\cite{Chavez2016}.

A major issue that needs to be addressed in future work concerns
a possible uncertainty in the initial mass function (IMF), or the possibility
that it may vary with redshift. Since the $L({\rm H}\beta$) - $\sigma$
relation correlates the ionizing flux produced by the massive stars to
the velocity field in the potential well created primarily by the gas
and lower mass stars, any systematic variation of the IMF affects the 
mass-to-light ratio and, therefore, the slope and zero point of the 
correlation function \cite{Chavez2014}. There is hope that at
least some of these uncertainties may be reduced further with
the help of upcoming observations using the K-band Multi Object 
Spectrograph at the Very Large Telescope. This survey should produce
a larger sample of high-$z$ HIIGx, providing much better and competitive 
constraints on the cosmological parameters and the comparison of
competing models \cite{Terlevich2015}.

\acknowledgments
We are grateful to the anonymous referee for pointing out several
improvements that have led to a better presentation of the results.
FM is also grateful to the Instituto de Astrof\'isica 
de Canarias in Tenerife and to Purple Mountain Observatory in Nanjing, China
for their hospitality while part of this research was carried out. FM is also 
grateful for partial support to the Chinese Academy of Sciences Visiting 
Professorships for Senior International Scientists under grant 2012T1J0011,
and to the Chinese State Administration of Foreign Experts Affairs under
grant GDJ20120491013.


\begin{thebibliography}{99}

\bibitem[1]{Melnick1987} Melnick, J., Moles, M., Terlevich, R. \& Garcia-Pelayo, J.-M., ``Giant H II regions as distance indicators. I - Relations between 
global parameters for the local calibrators," MNRAS, {\bf 226}, 849 (1987)
\bibitem[2]{Searle1972} Searle, L. \& Sargent, W.L.W., ``Inferences from the Composition of Two Dwarf Blue Galaxies," ApJ, {\bf 173}, 25 (1972)
\bibitem[3]{Bergeron1977} Bergeron, J., ``Characteristics of the blue stars in the dwarf galaxies I ZW 18 and II ZW 40," ApJ, {\bf 211}, 62 (1977)
\bibitem[4]{Terlevich1981} Terlevich, R. \& Melnick, J., ``The dynamics and chemical composition of giant extragalactic H II regions," MNRAS, {\bf 195}, 839 (1981)
\bibitem[5]{Kunth2000} Kunth, D. \& {\"O}stlin, G., ``The most metal-poor galaxies," A\&ARv, {\bf 10}, 1 (2000)
\bibitem[6]{Melnick2000} Melnick, J., Terlevich, R. \& Terlevich, E., ``Hii galaxies as deep cosmological probes," MNRAS, {\bf 311}, 629 (2000)
\bibitem[7]{Siegel2005} Siegel, E.~R., Guzm{\'a}n, R., Gallego, J.~P., Ordu{\~n}a-L{\'o}pez, M. \& Rodr{\'{\i}}guez-Hidalgo, P., ``Towards a precision cosmology 
from starburst galaxies at $z > 2$," MNRAS, {\bf 356}, 1117 (2005)
\bibitem[8]{Melnick1988} Melnick, J., Terlevich, R. \& Moles, M., ``Giant H II regions as distance indicators. II - Application to H II galaxies and the value of 
the Hubble constant," MNRAS, {\bf 235}, 297 (1988)
\bibitem[9]{Fuentes2000} Fuentes-Masip, O., Mu{\~n}oz-Tu{\~n}{\'o}n, C., Casta{\~n}eda, H.~O. \& Tenorio-Tagle, G., 
``On the Size and Luminosity versus Velocity Dispersion Correlations from the Giant H II Regions in the Irregular Galaxy NGC 4449," AJ, {\bf 120}, 752 (2000)
\bibitem[10]{Bosch2002} Bosch, G., Terlevich, E. \& Terlevich, R., ``Narrow-band CCD photometry of giant H II regions," MNRAS, {\bf 329}, 481 (2002)
\bibitem[11]{Telles2003} Telles, E., ``HII Galaxies," ASPC, {\bf 297}, 143 (2003)
\bibitem[12]{Bordalo2011} Bordalo, V. \& Telles, E., ``The $L-\sigma$  Relation of Local H II Galaxies," ApJ, {\bf 735}, 52 (2011)
\bibitem[13]{Plionis2011} Plionis, M., Terlevich, R., Basilakos, S., Bresolin, F., Terlevich, E., Melnick, J. \& Chavez, R., ``A strategy to measure the dark 
energy equation of state using the H II galaxy Hubble function and X-ray active galactic nuclei clustering: preliminary results," MNRAS, {\bf 416}, 2981 (2011)
\bibitem[14]{Mania2012} Mania, D. \& Ratra, B., ``Constraints on dark energy from H II starburst galaxy apparent magnitude versus redshift data,"
PhLB, {\bf 715}, 9 (2012)
\bibitem[15]{Chavez2012} Ch{\'a}vez, R., Terlevich, E., Terlevich, R., Plionis, M., Bresolin, F., Basilakos, S. \& Melnick, J., 
``Determining the Hubble constant using giant extragalactic H II regions and H II galaxies," MNRAS Lett, {\bf 425}, L56 (2012)
\bibitem[16]{Chavez2014} Ch{\'a}vez, R., Terlevich, R., Terlevich, E., Bresolin, F., Melnick, J., Plionis, M. \& Basilakos, S., 
``The $L-\sigma$  relation for massive bursts of star formation," MNRAS, {\bf 442}, 3565 (2014)
\bibitem[17]{Terlevich2015} Terlevich, R., Terlevich, E., Melnick, J., Ch{\'a}vez, R., Plionis, M., Bresolin, F. \& Basilakos, S., 
``On the road to precision cosmology with high-redshift H II galaxies," MNRAS, {\bf 451}, 3001 (2015)
\bibitem[18]{Wei2016} Wei, J.-J., Wu, X.-F. \& Melia, F., ``The HII Galaxy Hubble Diagram Strongly Favors $R_{\rm h}=ct$ Over $\Lambda$CDM," 
MNRAS, {\bf 463}, 1144 (2016)
\bibitem[19]{Melia2003} Melia, F., ``The Edge of Infinity: Supermassive Black Holes in the Universe" (New York: Cambridge  University Press), 158 (2003)
\bibitem[20]{Melia2007} Melia, F., ``The Cosmic Horizon," MNRAS, {\bf 382}, 1917 (2007)
\bibitem[21]{Melia2013a} Melia, F., ``The $R_{\rm h}=ct$ Universe Without Inflation," A\&A, {\bf 553}, A76 (2013)
\bibitem[22]{Melia2016} Melia, F., ``Physical Basis for the Symmetries in the Friedmann-Robertson-Walker Metric," Frontiers of Physics, {\bf 11}, 119801 (2016)
\bibitem[23]{Melia2017a} Melia, F., ``The Linear Growth of Structure in the $R_{\bf h}=ct$ Universe," MNRAS, {\bf 464}, 1966 (2017)
\bibitem[24]{MeliaAbdelqader2009} Melia, F. \& Abdelqader, M., ``The Cosmological Spacetime,"  {\bf IJMP-D}, 18, 1889 (2009)
\bibitem[25]{MeliaShevchuk2012} Melia, F. \& Shevchuk, A., ``The $R_{\rm h}=ct$ Universe," MNRAS, {\bf 419}, 2579 (2012)
\bibitem[26]{Ade2014} Ade, P.A.R. et al., ``Planck 2013 results. XXIII. Isotropy and statistics of the CMB," A\&A, {\bf 571}, id. A23 (2014)
\bibitem[27]{Melia2012} Melia, F., ``Analysis of the Union2.1 SN Sample with the $R_{\rm h}=ct$ Universe," AJ, {\bf 144}, 110 (2012)
\bibitem[28]{Wei2015b} Wei, J.-J., Wu, X.-F., Melia, F., Maier, R.~S., ``A Comparative Analysis of the Supernova Legacy Survey Sample with LCDM and 
the $R_{\rm h}=ct$ Universe," AJ, {\bf 149}, 102 (2015)
\bibitem[29]{Perlmutter1998} Perlmutter, S. et al., ``Discovery of a supernova explosion at half the age of the universe," Nature, {\bf 391}, 51 (1998)
\bibitem[30]{Riess1998} Riess, A.~G. et al., ``Observational Evidence from Supernovae for an Accelerating Universe and a Cosmological Constant,"
AJ, {\bf 116}, 1009 (1998)
\bibitem[31]{Schmidt1998} Schmidt, B.~P. et al., ``The High-Z Supernova Search: Measuring Cosmic Deceleration and Global Curvature of the Universe 
Using Type IA Supernovae," ApJ, {\bf 507}, 46 (1998)
\bibitem[32]{Wei2013} Wei, J.-J., Wu, X. \& Melia, F., ``The Gamma-ray Burst Hubble Diagram and its Implications for Cosmology," ApJ, {\bf 772}, 43 (2013)
\bibitem[33]{Dai2004} Dai, Z.~G., Liang E.~W. \& Xu, D., ``Constraining $\Omega_m$ and Dark Energy with Gamma-Ray Bursts," ApJ Lett, {\bf 612}, L101 (2004)
\bibitem[34]{Ghirlanda2004} Ghirlanda, G., Ghisellini, G., Lazzati, D. \& Firmani, C., ``Gamma-Ray Bursts: New Rulers to Measure the Universe," 
ApJ Lett, {\bf 613}, L13 (2004)
\bibitem[35]{Melia2013b} Melia, F., ``High-$z$ Quasars in the $R_{\rm h}=ct$ Universe," ApJ, {\bf 764}, 72 (2013)
\bibitem[36]{Melia2014} Melia, F., ``The High-z Quasar Hubble Diagram," JCAP, {\bf 01}, 027 (2014)
\bibitem[37]{MeliaMcClintock2015} Melia, F. \& McClintock, T.~M., ``A Test of Cosmological Models Using High-$z$ Measurements of $H(z)$ ," AJ, {\bf 150}, 119 (2015)
\bibitem[38]{Kauffmann2000} Kauffmann, G. \& Haehnelt, M., ``A unified model for the evolution of galaxies and quasars," MNRAS, {\bf 311}, 576 (2000)
\bibitem[39]{Wyithe2003} Wyithe, J.S.B. \& Loeb A., ``Self-regulated Growth of Supermassive Black Holes in Galaxies as the Origin of the Optical and 
X-Ray Luminosity Functions of Quasars," ApJ, {\bf 595}, 614 (2003)
\bibitem[40]{MeliaMaier2013} Melia F. \& Maier, R., ``Cosmic Chronometers in the $R_{\rm h}=ct$ Universe," MNRAS, {\bf 432}, 2669 (2013)
\bibitem[41]{Jimenez2002} Jimenez, R. \& Loeb, A., ``Constraining Cosmological Parameters Based on Relative Galaxy Ages," ApJ, {\bf 573}, 37 (2002)
\bibitem[42]{Simon2005} Simon, J., Verde, L. \& Jimenez, R., ``Constraints on the redshift dependence of the dark energy potential," PhRvD, {\bf 71}, 123001 (2005)
\bibitem[43]{Wei2015c} Wei, J.-J., Wu, X.-F., Melia, F., Wang, F.-Y., Yu, H., ``Age-Redshift Relationship of Old Passive Galaxies," AJ, {\bf 150}, 35 (2015)
\bibitem[44]{Alcaniz1999} Alcaniz, J.~S. \& Lima, J.A.S., ``New Limits on $\Omega_\Lambda$ and $\Omega_m$ from Old Galaxies at High Redshift," 
ApJ, {\bf 521}, L87 (1999)
\bibitem[45]{Lima2000} Lima, J.A.S. \& Alcaniz, J.~S., ``Constraining the cosmic equation of state from old galaxies at high redshift," MNRAS, {\bf 317}, 893 (2000)
\bibitem[46]{Wei2014} Wei, J.-J., Wu X.-F., Melia F., ``A Comparison of Cosmological Models Using Time Delay Lenses," ApJ, {\bf 788}, id. 190 (2014)
\bibitem[47]{Melia2015} Melia, F., Wei, J.-J. \& Wu, X.-F., ``A Comparison of Cosmological Models Using Strong Gravitational Lensing Galaxies," AJ, {\bf 149}, 2 (2015)
\bibitem[48]{YennapureddyMelia2017} Yennapureddy, M. K. \& Melia, F., ``Cosmological Tests with Strong Gravitational Lenses using Gaussian Processes," 
AJ, submitted (2017)
 \bibitem[49]{Inserra2014} Inserra, C. \& Smartt, S.~J., ``Superluminous Supernovae as Standardizable Candles and High-redshift Distance Probes," 
ApJ, {\bf 796}, 87 (2014)
\bibitem[50]{Wei2015a} Wei, J.-J., Wu X.-F. \& Melia F., ``Testing Cosmological Models with Type Ic Super Luminous Supernovae," AJ, {\bf 149}, 165 (2015)
\bibitem[51]{Seikel2012}Seikel, M., Clarkson, C. \& Smith, M., ``Reconstruction of dark energy and expansion dynamics using Gaussian processes," 
JCAP, {\bf 06}, 036S (2012)
\bibitem[52]{DeLong1988}DeLong, E. R., DeLong, D. M. and Clarke-Pearson, D. L., ``Comparing the Areas under Two or More Correlated Receiver Operating Characteristic Curves: A Nonparametric Approach," Biometrics, {\bf 44}, 837 (1988)
\bibitem[53]{Bilicki2012}Bilicki, M. and Seikel, M., ``We do not live in the $R_{\rm h}=ct$ Universe," MNRAS, {\bf 425}, 1664 (2012)
\bibitem[54]{Wei2017} Wei, J.-J., Melia, F. and Wu, X., ``Impact of a Locally Measured H0 on the Interpretation 
of Cosmic-chronometer Data," ApJ, {\bf 835}, id.270 (2017)
\bibitem[55]{Chavez2016} Ch{\'a}vez, R., Plionis, M., Basilakos, S., Terlevich, R., Terlevich, E., Melnick, J., Bresolin, F. \&
Gonz{\'a}lez-Mor{\'a}n A.~L., ``Constraining the dark energy equation of state with H II galaxies," MNRAS, {\bf 462}, 2431 (2016)

\end{thebibliography}
\end{document}